\documentclass[universe,article,submit,pdftex,moreauthors]{Definitions/mdpi} 

\firstpage{1} 
\pubvolume{1}
\issuenum{1}
\articlenumber{0}
\pubyear{2022}
\copyrightyear{2022}
\datereceived{} 
\dateaccepted{} 
\datepublished{} 
\hreflink{https://doi.org/} 

\pdfoutput=1

\Title{Astroparticle Constraints from the Cosmic Star Formation Rate Density at High Redshift: Current Status and Forecasts for JWST}

\TitleCitation{Astroparticle constraints from the cosmic SFR density at high-$z$}

\Author{Giovanni Gandolfi $^{1,2,3}$\orcidA{}, Andrea Lapi $^{1,2,3,4}$\orcidB{}, Tommaso Ronconi $^{1,2}$\orcidC{} and Luigi Danese $^{1,2}$\orcidD}

\AuthorCitation{Gandolfi, G., Lapi, A.; Ronconi, T.; Danese, L.}

\address{$^{1}$ \quad Scuola Internazionale Superiore Studi Avanzati (SISSA), Physics Area, Via Bonomea 265, 34136 Trieste, Italy; ggandolf@sissa.it (G.G.); lapi@sissa.it (A.L.); tronconi@sissa.it (T.R.); danese@sissa.it (L.D.)\\
$^{2}$ \quad Institute for Fundamental Physics of the Universe (IFPU), Via Beirut 2, 34014 Trieste, Italy\\
$^{3}$ \quad Istituto Nazionale Fisica Nucleare (INFN), Sezione di Trieste, Via Valerio 2, 34127 Trieste,  Italy\\
$^{4}$ \quad Istituto di Radio-Astronomia (IRA-INAF), Via Gobetti 101, 40129 Bologna, Italy\\}

\corres{Correspondence: ggandolf@sissa.it}

\abstract{We exploit the recent determination of the cosmic star formation rate (SFR) density at high redshifts $z\gtrsim 4$ to derive astroparticle constraints on three common dark matter (DM) scenarios alternative to standard cold dark matter (CDM): warm dark matter (WDM), fuzzy dark matter ($\psi$DM) and self-interacting dark matter (SIDM). Our analysis relies on the ultraviolet (UV) luminosity functions measured from blank-field surveys by the Hubble Space Telescope out to $z\lesssim 10$ and down to UV magnitudes $M_{\rm UV}\lesssim -17$. We extrapolate these to fainter yet unexplored magnitude ranges, and perform abundance matching with the halo mass functions in a given DM scenario, so obtaining a redshift-dependent relationship between the UV magnitude and the halo mass. We then compute the cosmic SFR density by integrating the extrapolated UV luminosity functions down to a faint magnitude limit $M_{\rm UV}^{\rm lim}$, which is determined via the above abundance matching relationship by two free parameters: the minimum threshold halo mass $M_{\rm H}^{\rm GF}$ for galaxy formation, and the astroparticle quantity $X$ characterizing each DM scenario (namely, particle mass for WDM and $\psi$DM, and kinetic temperature at decoupling $T_X$ for SIDM). We perform Bayesian inference on such parameters via a MonteCarlo Markov Chain (MCMC) technique by comparing the cosmic SFR density from our approach to the current observational estimates at $z\gtrsim 4$, constraining the WDM particle mass to $m_X\approx 1.2^{+0.3\,(11.3)}_{-0.4\,(-0.5)}$ keV, the $\psi$DM particle mass to $m_X\approx 3.7^{+1.8\,(+12.9.3)}_{-0.4\,(-0.5)}\times 10^{-22}$ eV, and the SIDM temperature to $T_X\approx 0.21^{+0.04\,(+1.8)}_{-0.06\,(-0.07)}$ keV at $68\%$ ($95\%$) confidence level. Finally, we forecast how such constraints will be strengthened by upcoming refined estimates of the cosmic SFR density, if the early data on the UV luminosity function at $z\gtrsim 10$ from the James Webb Space Telescope (JWST) will be confirmed down to ultra-faint magnitudes.} 

\keyword{dark matter; galaxy formation} 

\begin{document}

\section{Introduction}\label{sec|Intro}

Astrophysical and cosmological probes have firmly established that baryons constitute only some $15\%$ of the total matter content in the Universe. The rest is in the form of `dark matter’ (DM), which interacts very weakly or negligibly with the baryons except via long-range gravitational forces.
Still, no firm detection of DM particles has been made so far, despite the big efforts carried on with colliders \cite{Mitsou13,Kahlhoefer17,Argyropoulo21} or with direct \cite{Aprile18,Bernabei20} and indirect \cite{Fermi15,Ackermann17,Albert17,Zornoza21} searches in the sky.

The standard lore envisages DM to be constituted by weakly interacting particles with masses of order GeV \cite{Bertone18}, that are non-relativistic at the epoch of decoupling (hence they are dubbed 'cold' dark matter or CDM) and feature negligible free-streaming velocities (i.e., they do not diffuse out of perturbations before collapse). As a consequence, bound CDM structures called halos grow sequentially in time and hierarchically in mass by stochastically merging together \cite{Frenk12,Lapi20}.

On cosmological scale the CDM hypothesis is remarkably consistent with the data \cite{Planck20}, while on (sub)galactic scales it faces some challenges. For example, with respect to the predictions of gravity-only $N$-body simulations,
the shape of the inner density profiles in DM-dominated dwarfs is too flat \cite{Navarro97,deBlok08}, and the number and dynamical properties of observed Milky Way satellites differ from those of subhalos \cite{Boylan12,Bullock17}. Moreover, the emergence of tight empirical relationships between properties of the dark and luminous components in disc-dominated galaxies, such as the universal core surface density or the radial acceleration relation \cite{Gentile09,McGaugh16}, seem to be indicative of a new dark sector and/or of non-gravitational coupling between DM particles and baryons. Although the above effects can in principle be explained in CDM by invoking physical processes causing transfer of energy and angular momentum from the baryons to DM particles, such as dynamical friction \cite{ElZant01,Tonini06} or feedback effects from stars and active galactic nuclei \cite{Pontzen14,Peirani17,Freundlich20}, a fine-tuning is required to explain in detail the current data.

This has triggered the consideration of alternative, and perhaps more fascinating, solutions that rely on non-standard particle candidates \cite{Bertone04,Kusenko09,Feng10,Adhikari17,Salucci21}. The most widespread scenarios in the literature, which are also relevant for the present work include: warm dark matter (WDM) particle with masses in the keV range \cite{Bode01,Lovell14}; fuzzy or particle-wave dark matter ($\psi$DM), constituted by ultralight axion-like particles with masses $\gtrsim 10^{-22}$ eV \cite{Hu00,Hui17}; self-interacting dark matter (SIDM) with masses in the range $\sim 10-250$ MeV, as required by the cross-section $\sigma_{XX}/m_X\sim 0.1-1$ cm$^2$ g$^{-1}$ estimated from clusters to galaxies \cite{Vogelsberger16,Tulin18}. As a consequence of free-streaming, quantum pressure effects, and/or dark-sector interaction, all these scenarios produce a matter power spectrum suppressed on small scales, fewer (sub)structures, and flatter inner density profiles within halos relative to CDM \cite{Schneider12,Dayal15,Schive16,Huo18,Menci18,Lovell20,Romanello21,Kulkarni22}. Indirect astrophysical constraints on the properties of such nonstandard DM scenarios have been obtained by investigating the Lyman-$\alpha$ forest \cite{Viel13,Irsic17wdm,Irsic17fdm,Villasenor22}, high-redshift galaxy counts \cite{Pacucci13,Menci16,Shirasaki21,Sabti22}, $\gamma$-ray bursts \cite{deSouza12,Lapi17}, cosmic reionization \cite{Barkana01,Lapi15,Dayal17,Carucci19,Lapi22}, gravitational lensing \cite{Vegetti18,Ritondale19}, integrated 21cm data \cite{Carucci15,Boyarsky19,Chatterjee19,Rudakovskyi20}, $\gamma$-ray emission \cite{Bringmann17,Grand22}, fossil records of the Local Group \cite{Weisz14,Weisz17}, dwarf galaxy profiles and scaling relations \cite{Calabrese16,Burkert20}, and Milky Way satellite galaxies \cite{Kennedy14,Horiuchi14,Lovell16,Nadler21,Newton21} or a combination of these \cite{Enzi21}.

The present paper will focus on the constraints to DM that can be derived from recent observations of the cosmic SFR density at high redshift $z\gtrsim 4$ (e.g., \cite{Oesch18,Bouwens21,Bouwens22,Harikane22}). This observable crucially depends on the number density of ultra-faint galaxies, which   
tend to live within small halos, and especially so at high redshifts. Thus their numbers can constrain the shape of the halo mass distribution and of the power spectrum at the low mass end, which is sensitive to the microscopic properties of the DM particles. With respect to other probes of DM exploited in the literature, 
the cosmic SFR density is a very basic astrophysical quantity that suffer less from observational, systematic and modeling uncertainties. 

More in detail, we build up an empirical model based on the UV luminosity functions measured from blank-field surveys by the Hubble Space Telescope out to $z\lesssim 10$ and UV magnitudes $M_{\rm UV}\lesssim -17$. We extrapolate these to fainter yet unexplored magnitudes, and perform abundance matching with the halo mass functions in a given DM scenario, so obtaining a redshift-dependent relationship between the UV magnitude and the halo mass. We then compute the cosmic SFR density by integrating the extrapolated UV luminosity function down to a faint magnitude limit $M_{\rm UV}^{\rm lim}$, which is determined via the above abundance matching relationship by two free parameters describing our astrophysical and astroparticle uncertainties: the minimum threshold halo mass $M_{\rm H}^{\rm GF}$ for galaxy formation, and a quantity $X$ specific to each DM scenario (e.g., WDM particle mass). We perform Bayesian inference on the two parameters $M_{\rm H}^{\rm GF}$ and $X$ via a standard MCMC technique by comparing the cosmic SFR density from our approach to the current observational estimates at $z\gtrsim 4$. Finally, we forecast how the constraints on these parameters will be strengthened by upcoming refined estimates of the cosmic SFR density at $z\gtrsim 10$, if the early data on the UV luminosity function from the James Webb Space Telescope (JWST) will be confirmed down to ultra-faint magnitudes.

The structure of the paper is straightforward: in Section \ref{sec|methods} we describe our methods and analysis; in Section \ref{sec|results} we present and discuss our results; in Section \ref{sec|summary} we summarize our findings and highlight future perspectives. Throughout the work, we adopt the standard, flat cosmology \cite{Planck20} with rounded parameter values: matter density $\Omega_M \approx 0.31$, baryon density $\Omega_b \approx 0.05$, Hubble constant $H_0 = 100\, \rm{h}$ km s$^{-1}$ Mpc$^{-1}$ with $h\approx 0.68$. A Chabrier \cite{Chabrier03} initial mass function (IMF) is assumed.

\section{Methods and analysis}\label{sec|methods}

\begin{figure}[H]
\includegraphics[width=0.8\textwidth]{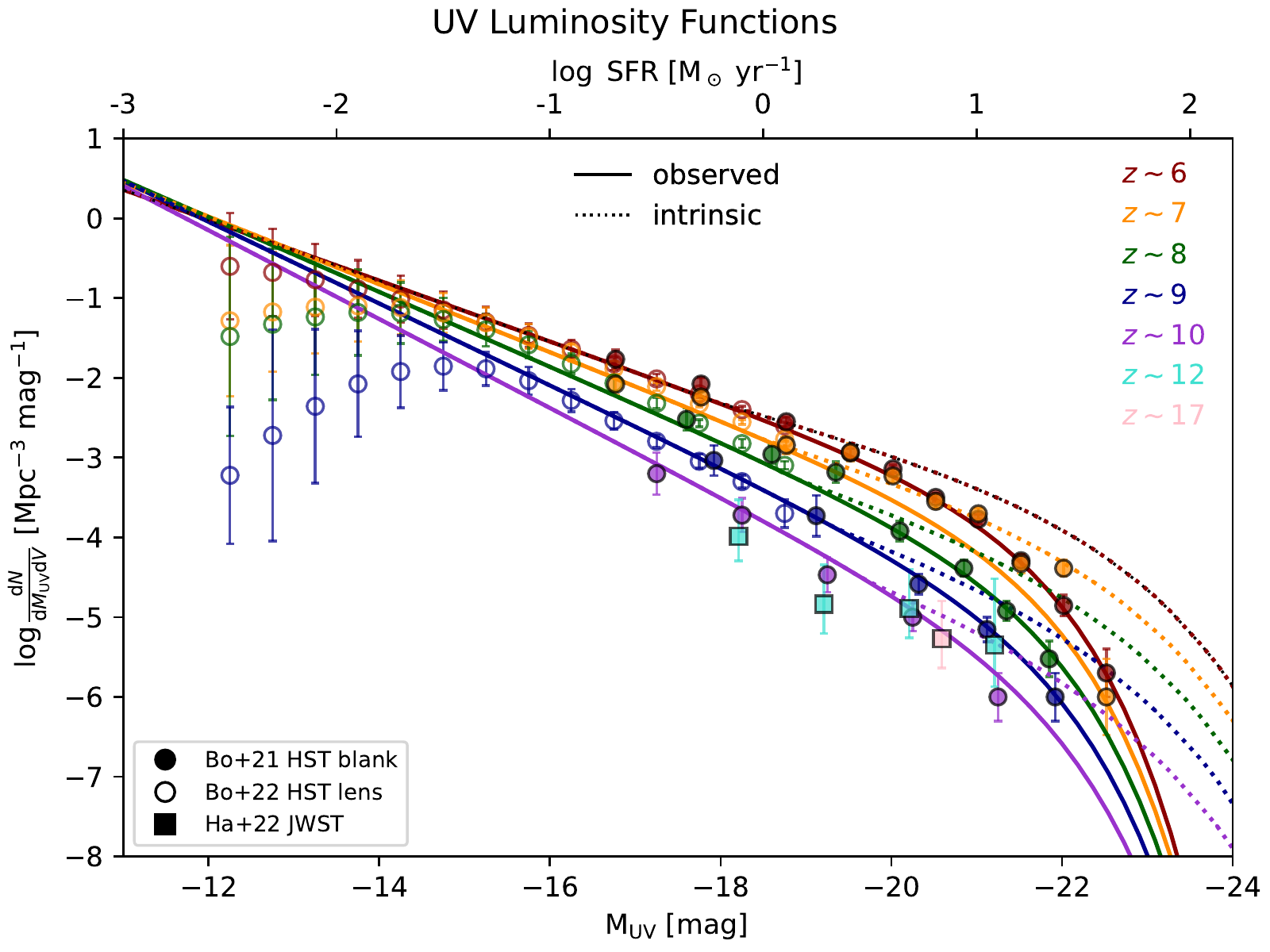}
\caption{The UV luminosity functions at redshifts $z\sim 6$ (red), $7$ (orange), $8$ (green), $9$ (blue), $10$ (magenta), $12$ (cyan) and $17$ (pink). Data points are from \cite{Bouwens21,Oesch18} (filled circles), \cite{Bouwens22} (empty circles), and \cite{Harikane22} (squares). Colored lines illustrate Schechter fits to the blank-field measurements from \cite{Bouwens21}: solid lines refer to the observed luminosity functions, while dotted lines to the intrinsic ones, after correction for dust extinction via the UV continuum slope according to the procedure by \cite{Bouwens14}.}\label{fig|UVLF}
\end{figure}

We start from the recent determination of the UV luminosity functions by \cite{Bouwens21,Oesch18} out to redshift $z\sim 10$ and UV magnitudes $M_{\rm UV}\lesssim -17$. In Figure~\ref{fig|UVLF} we illustrate the binned luminosity functions (filled circles) at $\approx 1600$ {\AA} in the relevant redshift range $z\sim 6-10$ (color-coded), together with the corresponding continuous Schechter function rendition (solid lines) in the form
\begin{equation}\label{eq|UVLF}
\cfrac{{\rm d}N}{{\rm d}M_{\rm UV}\,{\rm d}V} = \phi^\star\, \cfrac{\ln(10)}{2.5}\, 10^{-0.4\,(M_{\rm UV}-M_{\rm UV}^\star)\,(\alpha+1)}\times e^{-10^{-0.4\,(M_{\rm UV}-M_{\rm
UV}^\star)}}    
\end{equation}
We characterize the evolution with redshift of the parameters entering Eq. (\ref{eq|UVLF}) according to the expressions by \cite{Bouwens21,Bouwens22}. Toward high $z$, these yield a steepening faint end-slope $\alpha\approx -1.95-0.11\, (z-6)$, an approximately constant characteristic magnitude $M_{\rm UV}^\star\approx -21.04-0.05\, (z-6)$ and an appreciably decreasing normalization $\phi^\star\approx 3.8\times 10^{-4-0.35\,(z-6)-0.027\, (z-6)^2}$ Mpc$^{-3}$.  
In Figure~\ref{fig|UVLF}, we also report the intrinsic luminosity functions after correction for dust extinction (dotted lines), which have been computed exploiting the relation between extinction, the slope of the UV spectrum, and observed UV magnitude by \cite{Meurer99,Bouwens14}; the effects of dust extinction on the UV luminosity function are minor for $M_{\rm UV}\gtrsim -17$, and will be irrelevant for this work. The intrinsic UV luminosity can be related to the physical SFR of galaxies; in particular, for a Chabrier IMF, age $\gtrsim 10^8$ years, and appreciably sub-solar metallicity the relation $\log$ SFR [M$_\odot$ year$^{-1}$] $\approx -0.4\,(M_{\rm UV}+18.5)$ holds (see \cite{Kennicutt12,Madau14,Cai14,Robertson15,Finkelstein19}), and the related values are reported on the top axis in Figure \ref{fig|UVLF}.

In Figure~\ref{fig|UVLF} we also report two other sets of data. The first one (open circles) is from \cite{Bouwens22}, which have been able to estimate the luminosity function down to $M_{\rm UV}\approx -12.5$ by exploiting gravitational lensed galaxies in the Hubble Frontier Field clusters. However, the considerable statistical uncertainties related to the paucity of detected sources, and the possible systematics in the lensing reconstruction and completeness issues do not yet allow to draw firm conclusions on the shape of the luminosity function at such ultra-faint magnitudes. The second set of data (filled squares) involves the early results from JWST by \cite{Harikane22}, which have provided an estimate of the luminosity function at $z\gtrsim 12$, though with a rather low statistics. Interestingly, it seems that at $z\sim 12$ the shape of the luminosity function is roughly consistent with the lower redshift estimates, though its evolution in normalization considerably slows down; more data are needed to confirm such a trend, which could be very relevant for the astroparticle constraints of this work as we will show and forecast. 

From the intrinsic UV luminosity functions, the cosmic SFR density can be computed as
\begin{equation}\label{eq|cSFR}
\rho_{\rm SFR}(z)=\int_{-\infty}^{\min[M_{\rm UV}^{\rm obs},M_{\rm UV}^{\rm lim}]}{\rm d}M_{\rm UV}\, \frac{{\rm d}N}{{\rm d}M_{\rm UV}\,{\rm d}V}\, {\rm SFR}\;, 
\end{equation}
where $M_{\rm UV}^{\rm obs}$ is the faintest limit probed by observations (e.g., $M_{\rm UV}\approx -13$ for \cite{Bouwens22}, or $\approx -17$ for \cite{Harikane22}), and $M_{\rm UV}^{\rm lim}$ represents a limiting magnitude down to which the luminosity function is steeply increasing, i.e. we consider that the SFR density is negligible contributed by magnitudes $M_{\rm UV}\gtrsim M_{\rm UV}^{\rm lim}$ fainter than such a limit. The quantity $M_{\rm UV}^{\rm lim}$ is somewhat uncertain: as mentioned above, the most recent and stringent constraints are from the analysis by \cite{Bouwens22}, which rules out the presence of a turnover in the luminosity function brightward of $M_{\rm UV}\sim -15.5$. Actually the data by \cite{Bouwens22} seem to suggest a possible flattening of the luminosity function for
$M_{\rm UV}\gtrsim -15$, but the large errors and the systematic uncertainties due to the paucity of sources as well as incompleteness issues do not allow to make robust conclusions; thus in the following we will not try to model the detailed shape of any possible bending and use instead the extrapolation of the steep Schechter fits to the data by \cite{Bouwens21} with a sharp limit at $M_{\rm UV}^{\rm lim}$ \footnote{In fact, one can easily adopt a smooth bending of the luminosity function and set instead the upper limit of integration in Eq. \ref{eq|cSFR} just to $M_{\rm UV}^{\rm obs}$. E.g., \cite{Bouwens22} empirically suggest to multiply Eq. (\ref{eq|UVLF}) by a factor $10^{0.4\, (\alpha+1)/2 \times (M_{\rm UV}+16)^2/(M_{\rm UV}^{\rm lim}+16)}$ for $M_{\rm UV}\gtrsim -16$. We have checked that in the computation of the cosmic SFR this produces practically indistinguishable results with respect to our simple treatment.}.

The rationale is that at magnitudes fainter than $M_{\rm UV}^{\rm lim}$, the luminosity function flattens or even bends downwards because the galaxy formation process becomes inefficient and/or because the power spectrum is cut-off due to the microscopic nature of DM. Below we connect such a magnitude limit to two parameters describing these effects: a threshold halo mass $M_{\rm H}^{\rm GF}$ below which galaxy formation is hindered because of various processes, like photo-suppression by the intense UV background or inefficiency in atomic cooling by the low temperature and metallicity of small halos at high redshift (see \cite{Efstathiou92,Sobacchi13,Finkelstein19}); an astroparticle properties $X$ specific of a given DM scenario (e.g., WDM mass), that characterizes the suppression of the power spectrum at small scales.

\subsection{Halo mass function and abundance matching}\label{sec|AbMa}

We consider three common non-standard DM scenarios alternative to CDM: warm dark matter (WDM), fuzzy dark matter ($\psi$DM), and self-interacting dark matter (SIDM). In all these scenarios, the number of small-mass halos is reduced relative to CDM; this is best specified in terms of the halo mass function, namely the number density of halos per comoving volume and halo mass $M_{\rm H}$ bins, which can be conveniently written in terms of the CDM one as
\begin{equation}\label{eq|HMF}
\frac{{\rm d}N}{{\rm d}M_{\rm H}\, {\rm d}V}= \frac{{\rm d}N_{\rm CDM}}{{\rm d} M_{\rm H}\, {\rm d}V}\, \left[1+\left(\frac{M_{\rm H}^{\rm cut}}{M_{\rm H}}\right)^\beta\right]^{-\gamma}\;,
\end{equation}
where $\beta$ and $\gamma$ are shape parameters, and $M_{\rm H}^{\rm cut}$ is a cutoff halo mass. We compute the CDM halo mass function by exploiting the Python \texttt{COLOSSUS} package \cite{Diemer18} and the fitting formula by \cite{Tinker08} for virial masses. The parameters $(\beta,\gamma)$ in Eq. (\ref{eq|HMF}) are instead derived from fits to the outcomes of numerical simulations in the considered DM scenarios; the related values of the parameters, and the literature works from which these are taken (\cite{Schneider12,Schive16,Huo18}), are reported in Table \ref{tab|HMFpar}. We stress that for deriving robust constraints on different DM scenarios based on the halo mass function it is extremely important to rely on the results from detailed simulations (as done here), and not on semi-analytic derivations based on the excursion set formalism, whose outcomes on the shape of the mass function for masses $M_{\rm H}\lesssim M_{\rm H}^{\rm cut}$ are rather sensitive to several assumptions (e.g., the  filter function used in deriving the mass variance from the power spectrum, the mass-dependence in the collapse barrier, etc.; see \cite{Schneider13,Lapi15,May22}).

\begin{table}[H]
\caption{Parameters describing the ratio of the halo mass function for different DM scenarios relative to the standard CDM in terms of the expression $[1+(M_{\rm H}^{\rm cut}/M_{\rm H})^\beta]^{-\gamma}$, where $M_{\rm H}$ is the halo mass and $M_{\rm H}^{\rm cut}$ is a characteristic cutoff scale, see Section \ref{sec|AbMa} for details. The values of the parameters $(\beta,\gamma)$, extracted from fits to the outcomes of numerical simulations in the considered DM scenarios, are taken from the literature studies referenced in the last column.}\label{tab|HMFpar}
\newcolumntype{C}{>{\centering\arraybackslash}X}
\begin{tabularx}{0.6\textwidth}{CCCC}
\toprule
\textbf{Scenario} &\boldmath{ $\beta$} & \boldmath{$\gamma$ }& \textbf{Ref.}\\
\midrule
\\
WDM & $1.0$ & $1.16$ & \cite{Schneider12}\\
&&&\\
$\psi$DM & $1.1$ & $2.2$ & \cite{Schive16}\\
&&&\\
SIDM & $1.0$ & $1.34$ & \cite{Huo18}\\
&&&\\
\bottomrule
\end{tabularx}
\end{table}

As to the cutoff mass $M_{\rm H}^{\rm cut}$, in WDM it is determined by free-streaming effects \cite{Schneider12} and reads $M_{\rm H}^{\rm cut} \approx 1.9\times 10^{10}$ M$_\odot\, (m_X/{\rm keV})^{-3.33}$ in terms of the particle mass $m_X$. However, note that this cutoff (often referred to half-mode) mass is substantially larger by factors of a few $10^3$ than the free streaming mass, i.e., the mass related to the typical length-scale for diffusion of WDM particles out of primordial perturbations. In $\psi$DM, $M_{\rm H}^{\rm cut} \approx 1.6\times 10^{10}$ M$_\odot$ $\, (m_X/10^{-22}\, {\rm eV})^{-1.33}$ is related to the coherent behavior of the particles \cite{Schive16} with mass $m_X$. In the SIDM scenario, $M_{\rm H}^{\rm cut}\approx 7\times 10^7$ M$_\odot\, (T_X/{\rm keV})^{-3}$ can be linked to the visible sector temperature $T_X$ when kinetic decoupling of the DM particles takes place \cite{Huo18}. 

\begin{figure}[H]
\includegraphics[width=\textwidth]{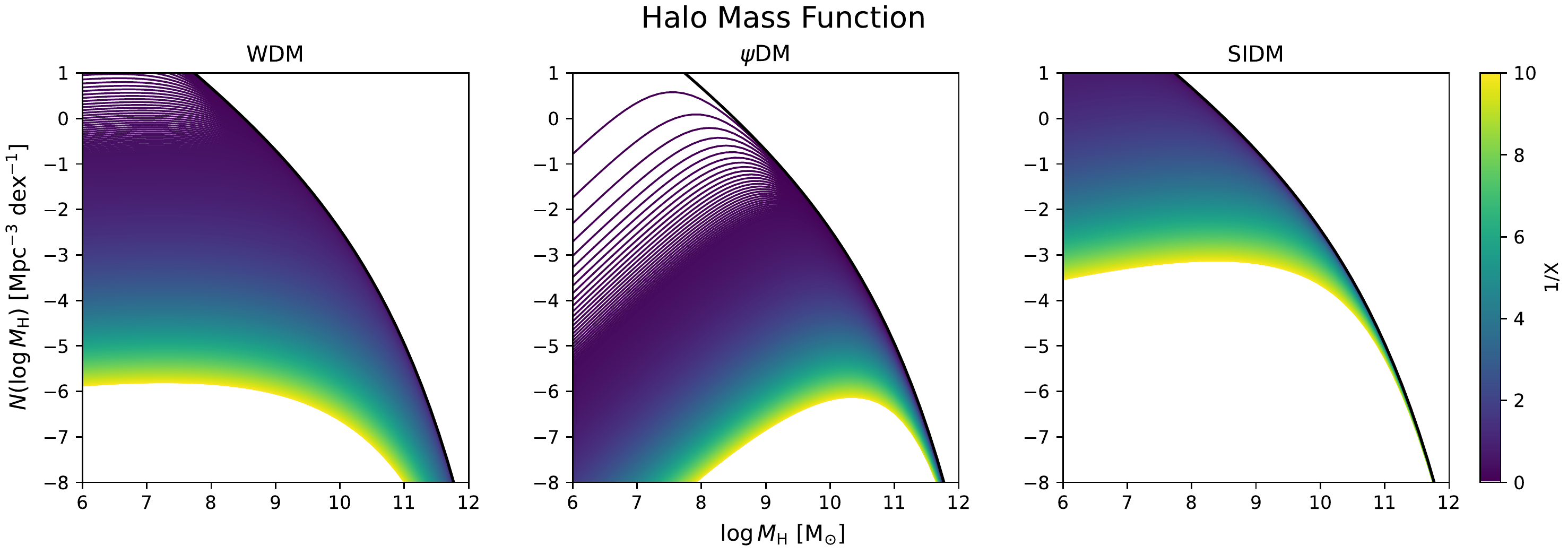}
\caption{Halo mass function at a reference redshift $z\approx 10$ in different DM scenarios: WDM (left panel), $\psi$DM (middle panel) and SIDM (right panel). The colorbar refers to values of keV$/m_X$ for WDM, $10^{-22}$ eV$/m_X$ for $\psi$DM and keV$/T_X$ for SIDM. In all panels the black line refers to the standard CDM scenario.}\label{fig|HMF}
\end{figure}

In Figure~\ref{fig|HMF}, we illustrate the halo mass functions in the different DM scenarios at a reference redshift $z\approx 10$, to highlight the dependence on the particle property. For example, focusing on WDM, it is seen that  the halo mass function progressively flattens with respect to that in standard CDM (black line); the deviation occurs at smaller halo masses for higher WDM particle masses $m_X$, so that the CDM behavior is recovered for $m_X\rightarrow \infty$. In the other DM scenarios, the behavior is similar, but the shape of the mass function past the low-mass end flattening can be appreciably different; e.g., in the $\psi$DM scenario the mass function is strongly suppressed for small masses and actually bends downward rather than flattening, implying a strong reduction or even an absence of low mass halos.

\begin{figure}[H]
\includegraphics[width=\textwidth]{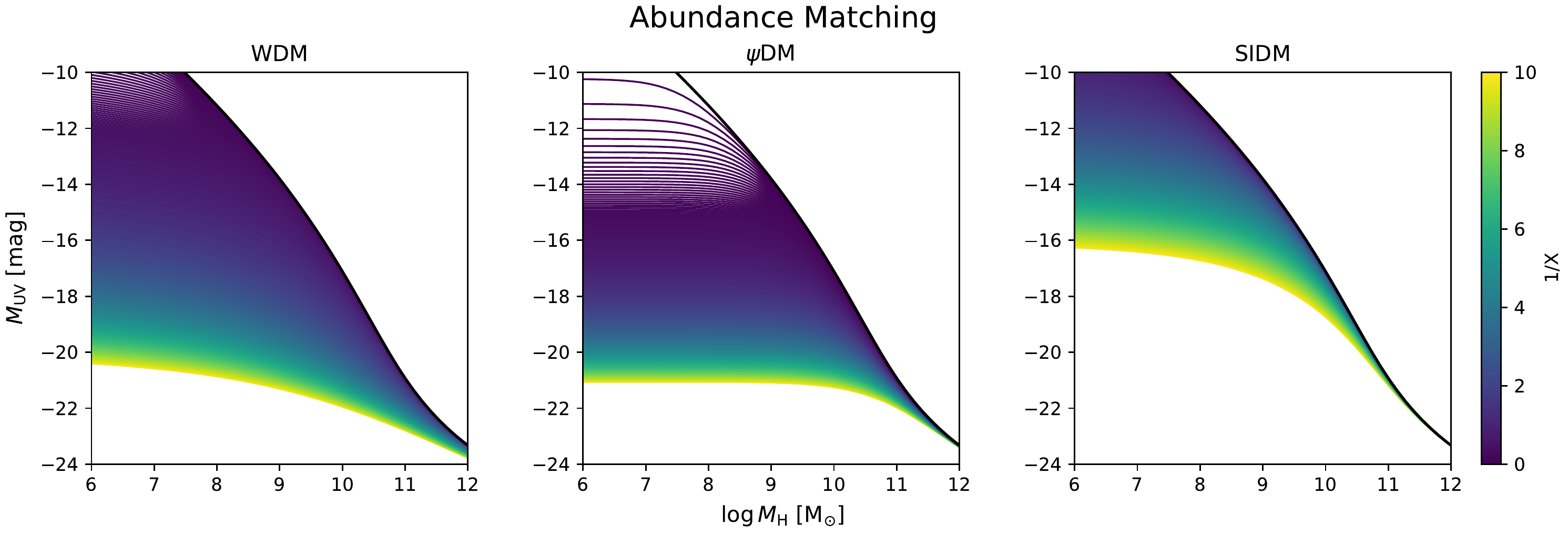}
\caption{Relationship between the UV magnitude $M_{\rm UV}$ and the halo mass $M_{\rm H}$ at a reference redshift $z\approx 10$, derived from the abundance matching of the observed UV luminosity function and the halo mass function (see text for details) in different DM scenarios: WDM (left panel), $\psi$DM (middle panel) and SIDM (right panel). The colorbar refers to values of keV$/m_X$ for WDM, $10^{-22}$ eV$/m_X$ for $\psi$DM and keV$/T_X$ for SIDM. In all panels the black line refers to the standard CDM scenario.}\label{fig|AbMa}
\end{figure}

We now look for a relationship between UV magnitude and halo masses via the standard abundance matching technique \cite{Aversa15,Moster18,Cristofari19,Behroozi20}, i.e., matching the cumulative number densities in galaxies and halos according to the expression
\begin{equation}\label{eq|abma}
\int_{M_{\rm H}}^{+\infty}{\rm d}M_{\rm H}'\;\frac{{\rm d}N}{{\rm d}M_{\rm H}'\, {\rm d}V}(M_{\rm H}',z|X) = \int_{-\infty}^{M_{\rm UV}}{\rm d}M_{\rm UV}'\;\frac{{\rm d}N}{{\rm d}M_{\rm UV}'\, {\rm d}V}(M_{\rm UV}',z)
\end{equation}
which implicitly defines a one-to-one monotonic relationship $M_{\rm UV}(M_{\rm H},z|X)$; here the quantity $X$ stands for the specific property of the DM scenario that determines the behavior of the mass function for $M_{\rm H}\lesssim M_{\rm H}^{\rm cut}$: particle mass $m_X$ in keV for WDM and in $10^{-22}$ eV for $\psi$DM, and kinetic temperature $T_X$ in keV for SIDM.  In Figure~\ref{fig|AbMa}, we show the outcome of this procedure at a reference redshift $z\approx 10$ in the different DM scenarios, highlighting its dependence on the particle property. Focusing on WDM as a representative case, it is seen that the $M_{\rm UV}(M_{\rm H},z|m_x)$ relation progressively flatten toward small $M_{\rm H} $with respect to the standard CDM case, and especially so for smaller $m_X$; at the other end, the relation becomes indistinguishable from that in CDM for particle masses $m_X\gtrsim$ some keVs. At a given particle mass, the relation $M_{\rm H}(M_{\rm UV},z|m_X)$ barely depends on redshift $z\gtrsim 6$, because the cosmic evolution of the UV luminosity function and the halo mass function mirror each other (see discussion by \cite{Bouwens21}). In the other DM scenarios, the behavior of the $M_{\rm UV}(M_{\rm H},z|X)$ relation is similar but its shape for small halo masses is appreciably different; e.g., in the $\psi$DM scenario, the relation flatten abruptly, reflecting the paucity of small halos in the mass function (see Figure \ref{fig|HMF}).

The rationale is now to compute the cosmic SFR density $\rho_{\rm SFR}(z)$ according to Eq. (\ref{eq|cSFR}) by integrating the luminosity function down to a magnitude limit $M_{\rm UV}^{\rm lim}(M_{\rm H}^{\rm GF},z|X)$ that depends on two parameters, namely the minimum halo mass for galaxy formation $M_{\rm H}^{\rm GF}$ and the astroparticle properties $X$ of a given DM scenario, and hence to estimate these quantities by comparing $\rho_{\rm SFR}(z)$ with the observational determinations.

\subsection{Bayesian analysis}\label{sec|Bayes}

The descriptions provided in the previous sections highlight that the limiting UV magnitude $M_{\rm UV}^{\rm lim}$ depends on two parameters: the limiting halo mass for galaxy formation $M_{\rm H}^{\rm GF}$, and a quantity $X$ specific to the DM scenario, that represent the particle mass $m_X$ in units of keV for WDM, the particle mass $m_X$ in units of $10^{-22}$ eV for $\psi$DM, and the temperature of kinetic decoupling $T_X$ in units of keV for SIDM. These two parameters are meant to effectively encompass a variety of effects determining $M_{\rm UV}^{\rm lim}$, related to the efficiency of the galaxy formation process in small halos, and to the suppression in the number of low-mass halos due to the microscopic nature of DM. An added value of the empirical approach pursued here, which relies on extrapolation of the observed UV luminosity functions down to $M_{\rm UV}^{\rm lim}$, is that no further parameter is needed to predict the cosmic SFR density (besides the underlying assumption of an IMF, that in any case marginally affects the astroparticle constraints, as shown by \cite{Lapi22}).

To estimate the two aforementioned parameters we adopt a Bayesian MCMC framework, numerically implemented via the Python package \texttt{emcee} \cite{Foreman13}. Since for large values of $X$ all the outcomes of the nonstandard scenario converge toward CDM, it is convenient to look for estimate of $1/X$ instead of $X$, so as to have a fitting parameter varying in a compact domain. We use a standard Gaussian likelihood $\mathcal{L}(\theta)\equiv -\sum_i \chi_i^2(\theta)/2$
where $\theta=\{M_{\rm H}^{\rm GF},1/X\}$ is the vector of parameters, and the summation is over different datasets; for the latter, the corresponding $\chi_i^2= \sum_j [\mathcal{M}(z_j,\theta)-\mathcal{D}(z_j)]^2/\sigma_{\mathcal{D}}^2(z_j)$ is obtained by comparing our empirical model expectations $\mathcal{M}(z_j,\theta)$ to the data $\mathcal{D}(z_j)$ with their uncertainties $\sigma_{\mathcal{D}}^2(z_j)$, summing over the different redshifts $z_j$ of the datapoints. Specifically, our overall data sample is constituted by robust measurements of the cosmic SFR density (see summary in Table \ref{tab|data}) from: UV luminosity function data from HST \cite{Bouwens22}, UV luminosity function early data from JWST \cite{Harikane22}, GRB counts data from Fermi \cite{Kistler09}, and (sub)mm luminosity function data from ALMA \cite{Gruppioni20}. In the computation of the cosmic SFR density we keep into account the minimum observational magnitude limit $M_{\rm UV}^{\rm obs}$ of the different datasets.

We adopt flat priors $\pi(\theta)$ on the parameters within the ranges $\log M_{\rm H}^{\rm GF} [M_\odot]\in [6,11]$, and $1/X\in [0,10]$. We then sample the posterior distribution $\mathcal{P}(\theta)\propto \mathcal{L}(\theta)\, \pi(\theta)$ by running $\texttt{emcee}$ with $10^4$ steps and $200$ walkers; each walker is initialized with a random position uniformly sampled from the (flat) priors. After checking the auto-correlation time, we remove the first $20\%$ of the flattened chain to ensure the burn-in; the typical acceptance fractions of the various runs are in the range $30-40\%$. 

\begin{table}[H]\tablesize{\footnotesize}
\caption{Overview of the estimate for the cosmic SFR density considered in the Bayesian analysis of this work. Values and uncertainties refer to $\log$ SFR $[M_\odot$  yr$^{-1}$]}\label{tab|data}
\newcolumntype{C}{>{\centering\arraybackslash}X}
\begin{tabularx}{\textwidth}{p{3cm}<{\centering} Cp{2.5cm}<{\centering}CC}
\toprule
\textbf{Data} & \textbf{Redshifts} & \textbf{Values} & \textbf{Uncertainties} & \textbf{Reference}\\
\midrule  
\\
& $\{3.8,4.9,5.9,$ & $\{-1.14,-1.4,-1.66,$ & $\{0.08,0.07,0.05,$ &  \\
UV LF [HST] & $6.8,7.9,8.9,$ & $-1.85,-2.05,-2.61,$ & $0.06,0.11,0.11,$ & \cite{Bouwens21,Oesch18,Bouwens22} \\ 
& $10.4\}$ & $-3.13\}$ & $0.35\}$ & \\
&&&\\
UV LF [JWST] & $\{\sim 9,\sim 12,$ & $\{-2.90,-3.61,$ & $\{0.17,0.27,$ & \cite{Harikane22} \\
& $\sim 17\}$ & $\lesssim -3.94\}$ & $0.31\}$ & \\
&&&\\
GRB counts [Fermi] & $\{4.49,5.49,$ & $\{-1.138,-1.423,$ & $\{0.184,0.289,$ & \cite{Kistler09} \\
& $6.49,7.74\}$ & $-1.262,-1.508\}$ & $0.359,0.517\}$ &  \\
&&&\\
(sub)mm LF [ALMA]  & $\{4.00,5.25\}$ & $\{-1.218,-1.252\}$ & $\{0.219,0.612\}$ & \cite{Gruppioni20} \\
\\
\bottomrule
\end{tabularx}
\end{table}

\section{Results and discussion}\label{sec|results}

\begin{figure}[H]
\includegraphics[width=0.7\textwidth]{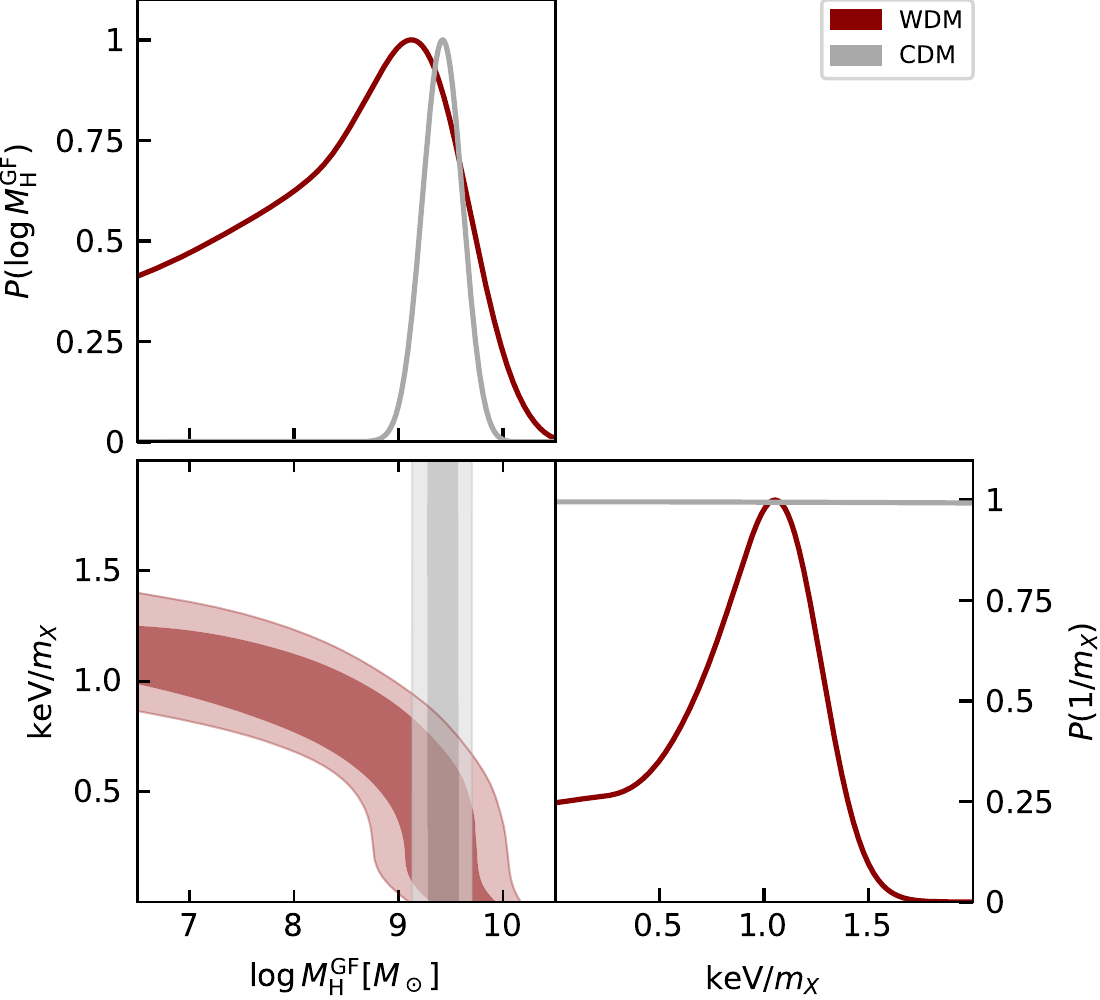}
\caption{MCMC posterior distributions in the WDM scenario (red contours/lines), for the threshold halo mass for galaxy formation $M_{\rm H}^{\rm GF}$, and the inverse of the DM particle's mass keV$/m_X$. For reference, the outcomes in the standard CDM scenario are also reported (grey contours/lines). The contours show $68\%$ and $95\%$ confidence intervals, and the marginalized distributions are in arbitrary units (normalized to $1$ at their maximum value).}\label{fig|WDM}
\end{figure}

\begin{figure}[H]
\includegraphics[width=0.7\textwidth]{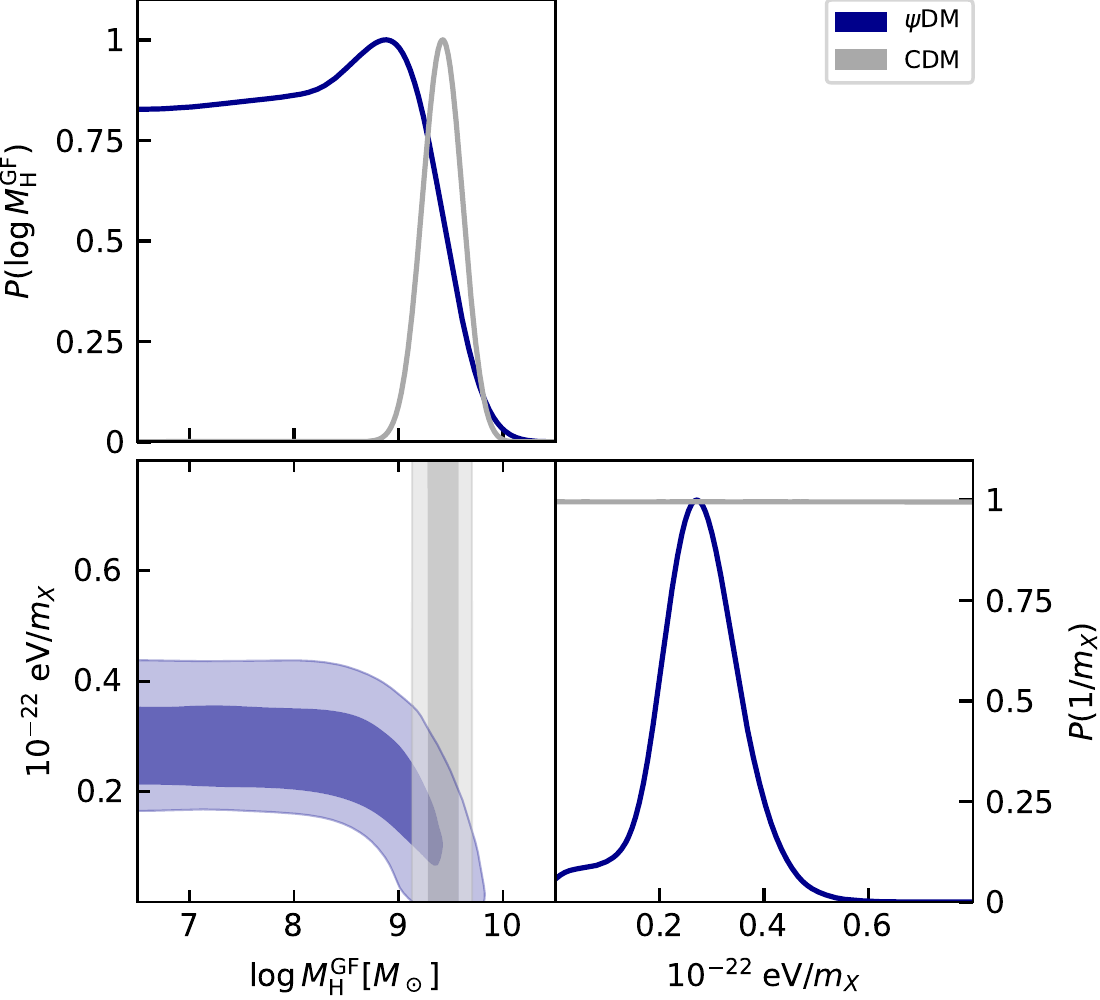}
\caption{MCMC posterior distributions in the $\psi$DM scenario (blue contours/lines), for the threshold halo mass for galaxy formation $M_{\rm H}^{\rm GF}$, and the inverse of the DM particle's mass $10^{-22}$ eV$/m_X$.
For reference, the outcomes in the standard CDM scenario are also reported (grey contours/lines). The contours show $68\%$ and $95\%$ confidence intervals, and the marginalized distributions are in arbitrary units (normalized to $1$ at their maximum value).}\label{fig|psiDM}
\end{figure}

\begin{figure}[H]
\includegraphics[width=0.7\textwidth]{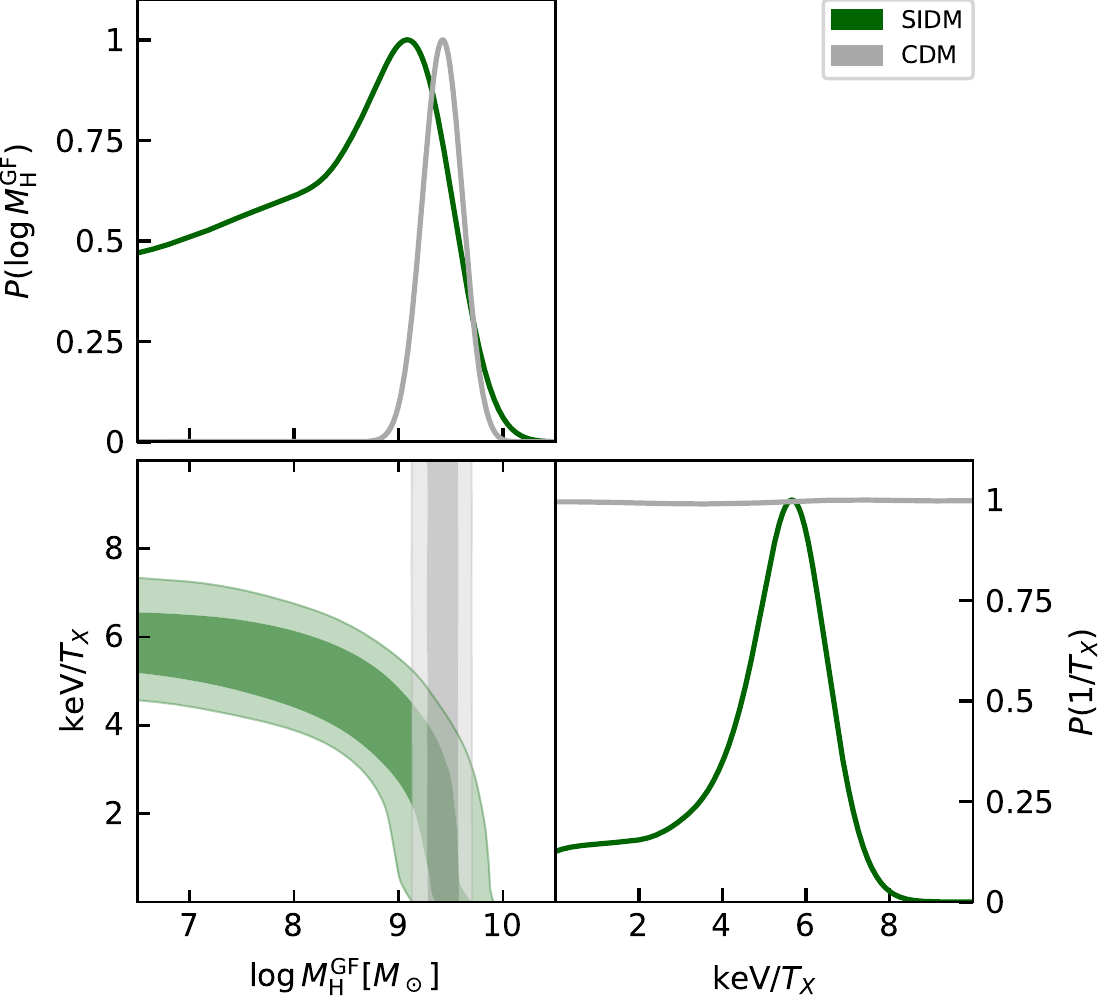}
\caption{MCMC posterior distributions in the SIDM scenario (green contours/lines), for the threshold halo mass for galaxy formation $M_{\rm H}^{\rm GF}$, and the inverse of the DM kinetic temperature at decoupling keV$/T_X$. For reference, the outcomes in the standard CDM scenario are also reported (grey contours/lines). The contours show $68\%$ and $95\%$ confidence intervals, and the marginalized distributions are in arbitrary units (normalized to $1$ at their maximum value).}\label{fig|SIDM}
\end{figure}

As a preliminary step, we analyze the data in the standard CDM scenario.  The result is shown by the grey contours/lines in Figure~\ref{fig|WDM}, \ref{fig|psiDM} and \ref{fig|SIDM}. By construction, in the CDM model the UV limiting magnitude $M_{\rm UV}^{\rm lim}$ depends only on the threshold minimum halo mass for galaxy formation. The marginalized constraint on the latter is found to be $\log M_{\rm H}^{GF} [M_\odot]\approx 9.4^{+0.2\,(+0.4)}_{-0.1\,(-0.4)}$, a value which is reasonably close to the photo-suppression mass expected by the intense UV background during reionization (e.g., \cite{Finkelstein19}). The corresponding limiting magnitude at $z\sim 10$ is around $M_{\rm UV}^{\rm lim}\approx -14.7$.

In the other DM scenarios, the situation is different, because the limiting UV magnitude can also depend on the DM astroparticle property $X$. 
The results for WDM are illustrated by the red lines/contours in Figure~\ref{fig|WDM}. It is seen that there is a clear degeneracy between the the WDM mass $m_X$ and the threshold halo mass $M_{\rm H}^{\rm GF}$ for galaxy formation, in that the same value of limiting UV magnitude $M_{\rm UV}^{\rm lim}$ can be obtained with smaller $M_{\rm H}^{\rm GF}$ and smaller $m_X$ (see Figure \ref{fig|AbMa}). This is because lowering $M_{\rm H}^{\rm GF}$ extends the halo mass function toward smaller masses and so allows more halos to be available for hosting galaxies, while decreasing $m_X$ progressively flattens the shape of the halo mass function so reducing the number of halos and offsetting the previous effect. Such a situation is possible if $m_X$ is not too low, otherwise the reduction in the number of halos is so drastic that cannot be compensated by reasonable values of $M_{\rm H}^{\rm GF}$; note the minimally acceptable $M_{\rm H}^{\rm GF}$ could be around $10^{7-8}\, M_\odot$, because below these masses atomic cooling becomes inefficient; a hard limit is set by minihalos of $10^6\, M_\odot$ where the first (pop-III) stars are thought to form.

The marginalized constraints for WDM turns out to be $\log M_{\rm H}^{GF} [M_\odot]\approx 7.6^{+2.2\,(+2.3)}_{-0.9\,(-3.3)}$ and $m_X\approx 1.2^{+0.3\,(11.3)}_{-0.4\,(-0.5)}$ keV, corresponding to a UV limiting magnitude $M_{\rm UV}^{\rm lim}\approx -13.3$. There is a clear peak in the posterior for the WDM mass around the keV scale, which is interesting because such a value has been often invoked to solve small-scale issues of CDM, like the missing satellites problems and the cusp-core controversy. However, larger values of $m_X\gtrsim $ a few keVs, that produce outcomes practically indistinguishable for CDM, are still well allowed (within $2\sigma$) by the current estimates of the cosmic SFR density. 

The situation for $\psi$DM and SIDM is somewhat similar to WDM. The main difference resides in the behavior of the halo mass function at small masses, which induces a different shape in the relationship between $M_{\rm H}$ and $M_{\rm UV}$, and in turn this affects the marginalized contraints .
In the $\psi$DM case, whose results are illustrated in Figure~\ref{fig|psiDM}, only an upper limit of the threshold halo mass for galaxy formation $\log M_{\rm H}^{GF} [M_\odot]< 7.9\, (<9.3)$ can be provided; however, the particle mass is constrained to $m_X\approx 3.7^{+1.8\,(+12.9.3)}_{-0.4\,(-0.5)}\times 10^{-22}$ eV, corresponding to a UV limiting magnitude $M_{\rm UV}^{\rm lim}\approx -14.6$ at $z\sim 10$.
In the SIDM scenario, whose results are illustrated in Figure~\ref{fig|psiDM}, the marginalized constraints reads $\log M_{\rm H}^{GF} [M_\odot]\approx 7.6^{+2.2\,(+2.3)}_{-1.1\,(-3.2)}$ and $T_X\approx 0.21^{+0.04\,(+1.8)}_{-0.06\,(-0.07)}$ keV, corresponding to a UV limiting magnitude $M_{\rm UV}^{\rm lim}\approx -13.7$ at $z\sim 10$. The overall marginalized constraints are summarized in Table \ref{tab|results}. 

\begin{table}[H]
\caption{Marginalized posterior estimates (mean, $68\%$ and $95\%$ confidence limits are reported) of the parameters from the MCMC analysis for the different DM scenarios considered in the main text (WDM, $\psi$DM and SIDM). Specifically, $M_{\rm H}^{\rm GF}$ is the threshold halo mass for galaxy formation, while the astroparticle quantity $X$ in the third column stands for: particle mass $m_X$ in keV for WDM, particle mass $m_X$ in $10^{-22}$ eV for $\psi$DM, and kinetic temperature $T_X$ in keV for SIDM; The last two columns refer to the value of the Bayes information criterion (BIC) and the Deviance information criterium (DIC) for model comparison, see Section \ref{sec|results}. The top half of the Table refers to the current constraints on the cosmic SFR density, while the bottom half to the forecasts for JWST observations extended down to UV magnitude $M_{\rm UV}\approx -13$, see Sect. \ref{sec|JWST} for details.}\label{tab|results}
\newcolumntype{C}{>{\centering\arraybackslash}X}
\begin{tabularx}{0.8\textwidth}{CCCCC}
\toprule
\textbf{Scenario} &\boldmath{ \textbf{$M_{\rm H}^{\rm GF}$}} & \boldmath{ \textbf{X}} & \textbf{BIC}& \textbf{DIC}\\
\midrule
\\
CDM  & $9.4^{+0.2\,(+0.4)}_{-0.1\,(-0.4)}$ & $-$ & $\approx 31$ & $\approx 13$\\
\\
WDM  & $7.6^{+2.2\,(+2.3)}_{-0.9\,(-3.3)}$ & $1.2^{+0.3\,(+11.3)}_{-0.4\,(-0.5)}$ & $\approx 33$ & $\approx 14$\\
\\
$\psi$DM & $<7.9\, (<9.3)$ &  $3.7^{+1.8\,(+12.9)}_{-0.9\,(-1.4)}$ & $\approx 33$ & $\approx 14$\\
\\
SIDM & $7.6^{+2.2\,(+2.3)}_{-1.1\,(-3.2)}$ & $0.21^{+0.04\,(+1.8)}_{-0.06\,(-0.07)}$ & $\approx 33$ & $\approx 14$\\
\\
\midrule
\\
CDM + JWST forecast & $<7.2\, (<8.5)$ & $-$ & $\approx 89$ & $\approx 130$\\
\\
WDM + JWST forecast & $<6.6\, (<8.2)$ & $>1.8\, (>1.2)$ & $\approx 87$ & $\approx 125$ \\
\\
$\psi$DM + JWST forecast & $6.2^{+1.3}_{-1.3}\, (<8.2)$ & $>17.3\, (>12)$ & $\approx 92$ & $\approx 135$ \\
\\
SIDM + JWST forecast & $<6.8\, (<8.3)$ & $>0.4\,(>0.3)$ & $\approx 89$ & $\approx 130$\\
\\
\bottomrule
\end{tabularx}
\end{table}

\begin{figure}[H]
\includegraphics[width=0.8\textwidth]{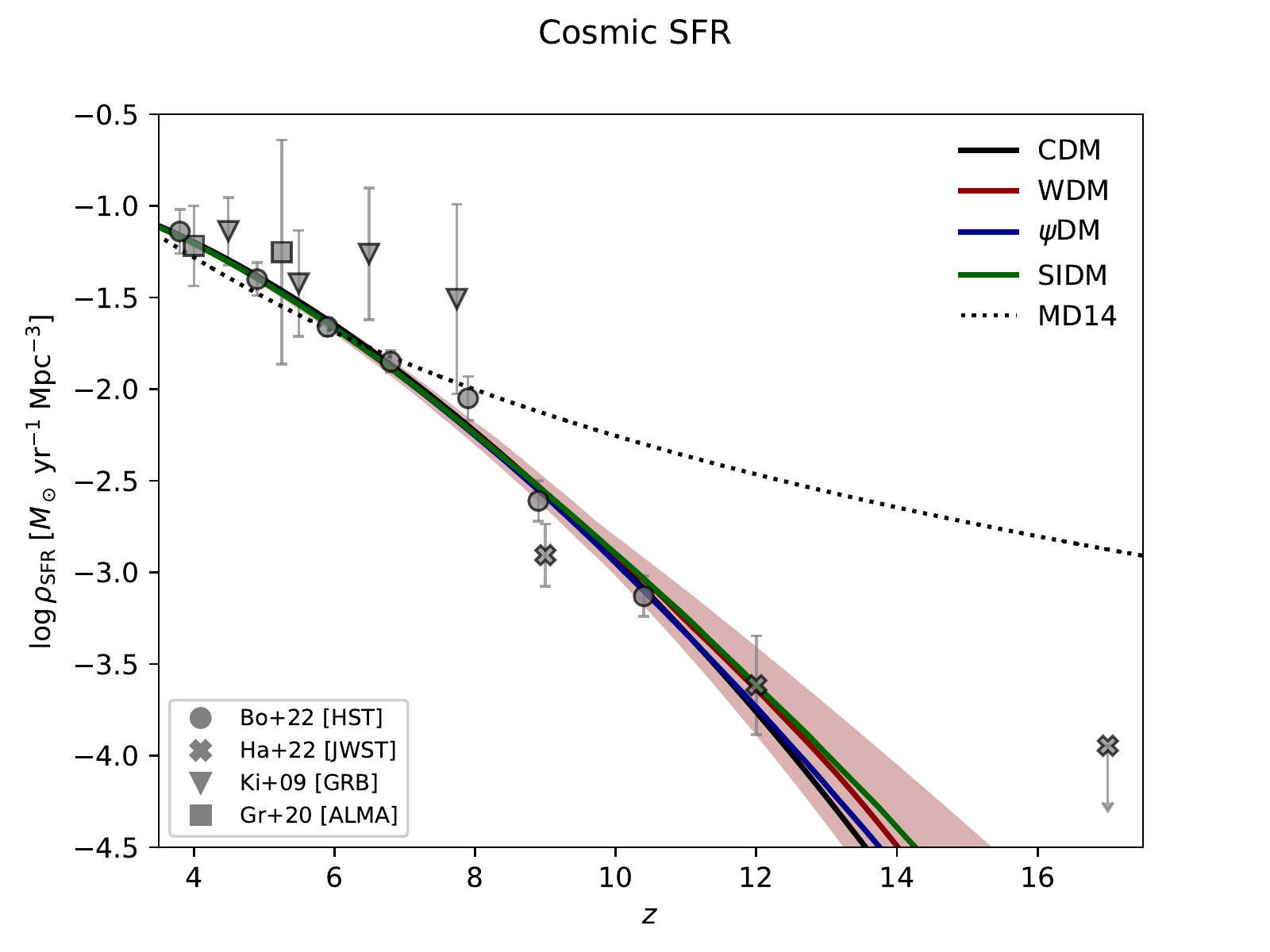}
\caption{The cosmic SFR density as a function of redshift. Data are from UV-HST (circles; \cite{Bouwens21,Oesch18}), UV-JWST (crosses; \cite{Harikane22}), GRB-Fermi (inverse triangles; \cite{Kistler09}) and (sub)mm-ALMA (squares; \cite{Gruppioni20}). Lines illustrate the best fits from the MCMC analysis in various DM scenarios:  CDM (black), WDM (red), $\psi$DM (blue), SIDM (green). The typical $2\sigma$ credible interval from sampling the posterior distribution is shown, for clarity, only in the WDM scenario, as a red shaded area. For reference, the dotted line is the classic fitting formula gauged at $z\lesssim 6$ by \cite{Madau14}.}\label{fig|SFR}
\end{figure}

In Figure \ref{fig|SFR} we illustrate the performance of our best-fits on the observed cosmic SFR density; all DM scenarios (colored lines) reproduce comparably well the available data. This is also highlighted by the $95\%$ credible interval from sampling the posterior distribution, which is shown only in the WDM case for clarity (red shaded area). In terms of projection on this observable, different DM scenarios are consistent with each other, approximately within $2\sigma$.

We can also attempt a quantitative model comparison analysis via the Bayes information criterion \cite{Schwarz78,Liddle04}, which is defined as 
BIC$\equiv -2\, \ln\mathcal{L_{\rm max}}+N_{\rm par}\, \ln N_{\rm data}$ in terms of the maximum likelihood estimate $\mathcal{L_{\rm max}}$, of the number of parameters $N_{\rm par}$, and the number of data points $N_{\rm data}$; the BIC comes from approximating the Bayes factor, which gives the posterior odds of one model against another, presuming that the models are equally favored a-priori. Another possibility, which may be less sensitive to priors is the Deviance information criterion \cite{Spiegelhalter02}, which is defined as DIC$\equiv -2\, \log\mathcal{L}(\bar\theta)+2\, p_D$ where the overbar denotes the mean and the effective number of parameters $p_D$ is estimated as $p_D\approx -2\,\overline{\log\mathcal{L}(\theta)}-2\, \log\mathcal{L}(\bar\theta)$. Note that what matters is only the relative value of the BIC or the DIC among different models; in particular, a difference larger than $10$ indicates robust evidence in favor of the model with the smaller value. The values of the BIC and the DIC (for the different DM scenarios) are reported in Table \ref{tab|results}, and do not suggest clear evidence in favor of one scenario over the others or over the standard CDM.

\subsection{Forecasts for JWST}\label{sec|JWST}

As mentioned in Sect. \ref{sec|methods} and shown in Figure \ref{fig|UVLF}, the early data from JWST at $z\sim 12$ seems to indicate a slowing down in the evolution of the UV luminosity function with respect to lower $z\lesssim 10$. The effect is evident also on the cosmic SFR density in Figure \ref{fig|SFR}, since the JWST data (crosses) at $z\sim 9-12$ are around the same value of the HST ones (circles), but the former refer to a UV luminosity function integrated down to $M_{\rm UV}^{\rm obs}\approx -17$ while the latter refer to $M_{\rm UV}^{\rm obs}\approx -13$.

Besides the possible issues related to systematics and completeness effects in the early JWST observations that will be hopefully cleared by future campaigns, one can ask the question: what if the JWST data will be confirmed and extended to ultra-faint magnitudes? To make a sound and conservative forecast on such a circumstance on the astroparticle constraints of this work, we proceed as follows. We scale up by $0.4$ dex the current SFR density estimate from JWST by \cite{Harikane22} at $z\gtrsim 9$, to reflect the same increase in $\rho_{\rm SFR}$ of the HST data by \cite{Bouwens22} when integrating the luminosity function from $M_{\rm UV}^{\rm obs}\approx -17$ to $M_{\rm UV}^{\rm obs}\approx -13$; we also assign a relative uncertainty to the JWST data comparable to that of the HST one by \cite{Bouwens22}.

\begin{figure}[H]
\begin{tabular}{@{}ccc}
\includegraphics[width=4.5cm]{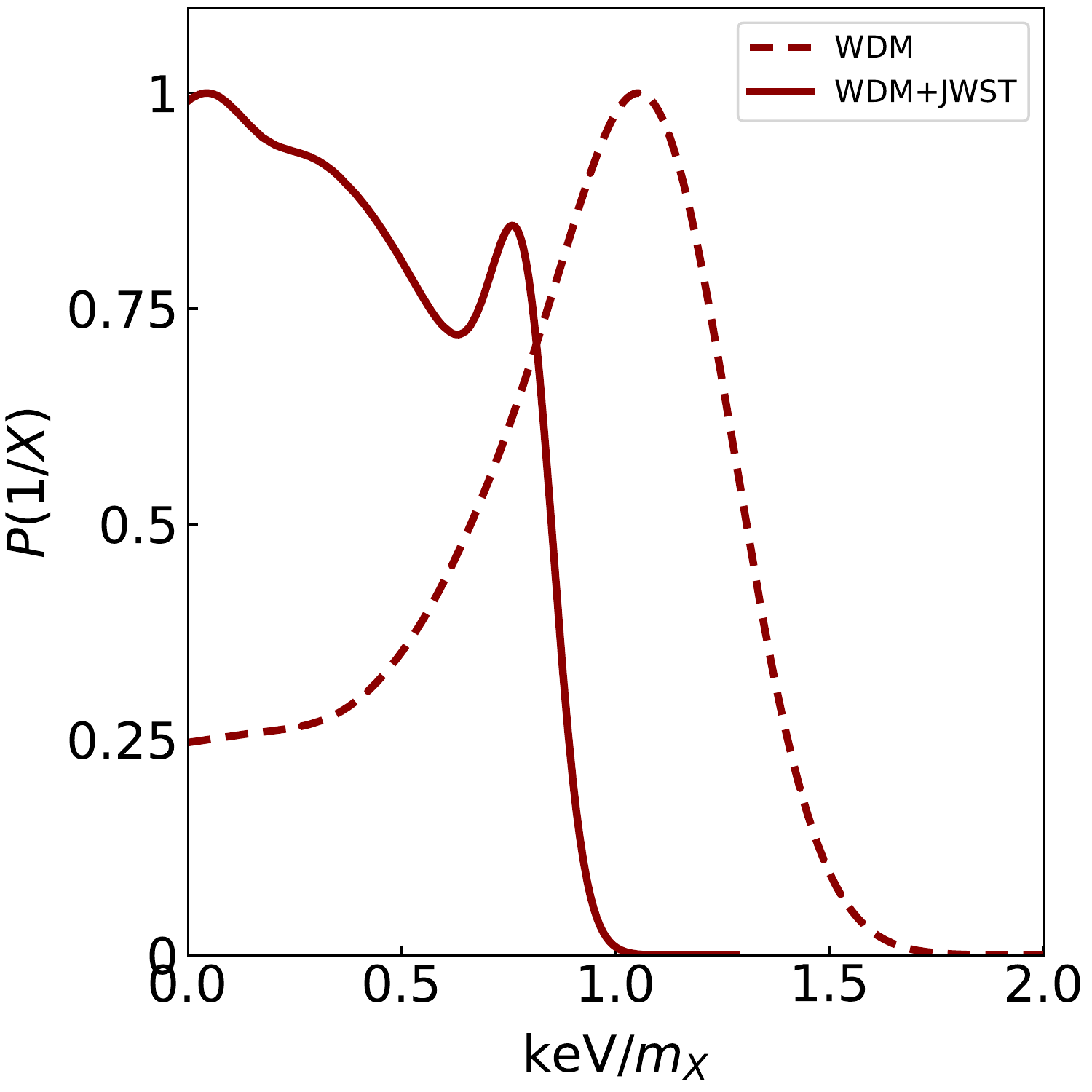}
\includegraphics[width=4.5cm]{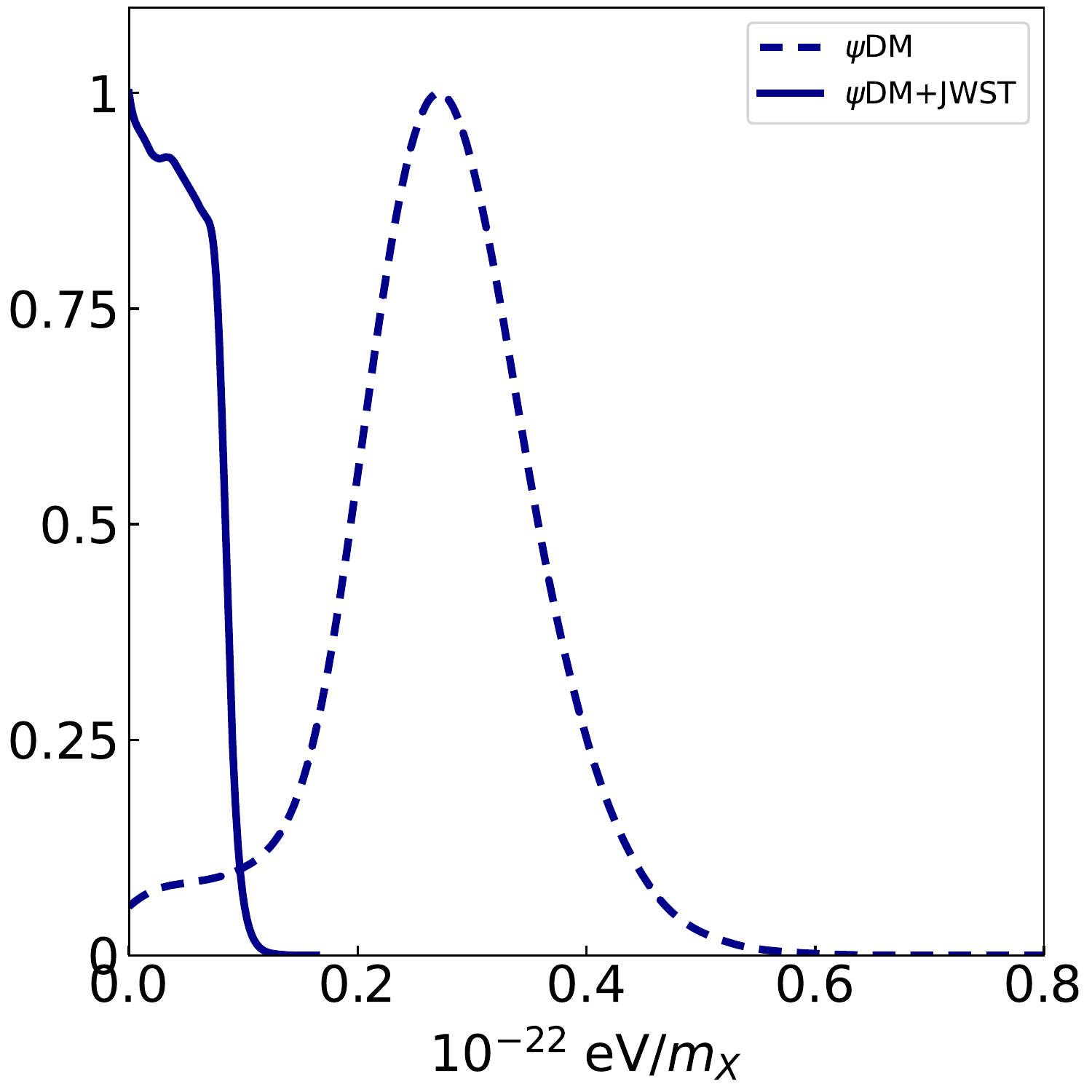}
\includegraphics[width=4.5cm]{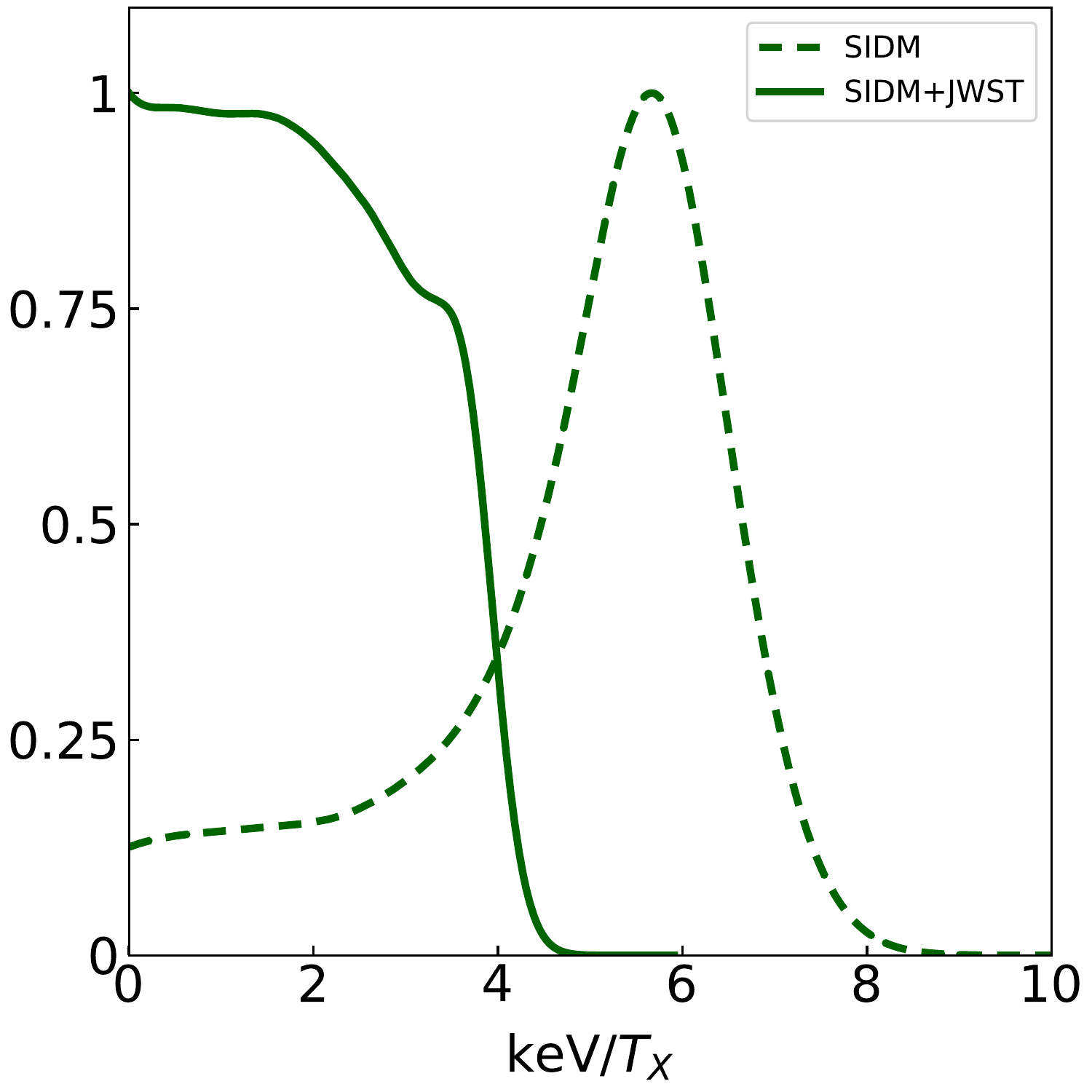}
\end{tabular}
\caption{Forecasts of the marginalized posteriors on the WDM mass (left panel), $\psi$DM mass (middle panel) and SIDM kinetic temperature at decoupling (right panel) based on prospective data at $z\gtrsim 10$ from JWST (solid lines; see text for details). For reference, the dashed lines illustrate the current constraints from Figs. \ref{fig|WDM}, \ref{fig|psiDM}, and \ref{fig|SIDM}. The marginalized distributions are in arbitrary units (normalized to $1$ at their maximum value).}\label{fig|forecast}
\end{figure}

In Figure \ref{fig|forecast} we illustrate the marginalized posteriors on the astroparticle quantities in the WDM, $\psi$DM and SIDM scenarios. Plainly, the appreciably higher values of the cosmic SFR density implied from the putative JWST data tend to go in tension with the suppression of the power spectrum at small scale in the non-CDM scenarios, erasing the bell-shaped posterior still allowed by the current data. As a consequence, rather stringent lower limits on the astroparticle quantities can be derived: WDM mass $m_X\gtrsim 1.8\, (1.2)$ keV, $\psi$DM mass $m_X\gtrsim 17.3\, (12)\times 10^{-22}$ eV, and SIDM kinetic temperature $T_X>0.4\, (0.3)$ keV. These lower bounds would be competitive with current literature constraints that tend to exclude part of the parameter space in non-CDM models (see references in Sect. \ref{sec|Intro}). Yet, the independent and basic nature of the cosmic SFR density observable, may provide constraints less affected by systematics and model-dependent interpretations.

Finally, note from Table \ref{tab|results} that the fit to the forecasted JWST data will require a quite low galaxy formation threshold $M_{\rm H}^{\rm GF}$ in CDM (and even more extreme values in the other scenarios); nonetheless, the upper bounds at $2\sigma$ are still consistent with the atomic cooling limit, so the forecasted JWST data should not present an insurmountable astrophysical challenge for CDM.  

\section{Summary}\label{sec|summary}

In this work, we have derived astroparticle constraints for different dark matter scenarios alternative to standard cold dark matter (CDM), namely warm dark matter (WDM), fuzzy dark matter ($\psi$DM) and self-interacting dark matter (SIDM), from recent determination of the cosmic star formation rate (SFR) density at high redshifts $z\gtrsim 4$. We have relied on the UV luminosity functions measured from blank-field surveys by the Hubble Space Telescope out to $z\lesssim 10$ and UV magnitudes $M_{\rm UV}\lesssim -17$. We have extrapolated these to fainter yet unexplored magnitudes, and performed abundance matching with the halo mass functions in a given DM scenario, so obtaining a redshift-dependent relationship between the UV magnitude and the halo mass. 

Then, we have computed the cosmic SFR density by integrating the extrapolated UV luminosity function down to a faint magnitude limit $M_{\rm UV}^{\rm lim}$, which is determined via the above abundance matching relationship by the minimum threshold halo mass $M_{\rm H}^{\rm GF}$ for galaxy formation, and by the astroparticle quantity $X$ specific to each DM scenario (e.g., WDM particle mass). 

Finally, we have performed Bayesian inference on the two parameters $M_{\rm H}^{\rm GF}$ and $X$ via a standard MCMC technique by comparing  the cosmic SFR density from our approach to the current observational estimates at $z\gtrsim 4$, so deriving definite astroparticle constraints: a WDM particle mass $m_X\approx 1.2^{+0.3\,(11.3)}_{-0.4\,(-0.5)}$ keV, a $\psi$DM particle mass $m_X\approx 3.7^{+1.8\,(+12.9)}_{-0.4\,(-0.5)}\times 10^{-22}$ eV, and a SIDM temperature at kinetic decoupling $T_X\approx 0.21^{+0.04\,(+1.8)}_{-0.06\,(-0.07)}$ keV at $68\%$ ($95\%$) confidence level. 

In addition, from the same analysis we have estimated that for CDM the minimum halo mass for galaxy formation is well constrained to $\log M_{\rm H}^{\rm GF} [M_\odot]\approx 9.4^{+0.2\, (+0.4)}_{-0.9\,(-0.4)}$, which is pleasingly close to the photo-suppression mass expected at high redshifts  due to the intense UV background. On the other hand, for non-CDM scenarios we have estimated a smaller $M_{\rm H}^{\rm GF}\lesssim 10^8\, M_\odot$, which is a value closer to the atomic cooling limit, although yet poorly constrained due to the degeneracy with the astroparticle property.

In a future perspective, we have forecasted how such constraints will be strengthened if the early data on the UV luminosity function at $z\gtrsim 10$ from the James Webb Space Telescope (JWST) will be confirmed and extended to ultra-faint magnitudes; these would imply upper limits on the WDM mass $m_X\gtrsim 1.8\, (\gtrsim 1.2)$ keV, on the $\psi$DM mass of $m_X\gtrsim 17.3\, (\gtrsim 12)\times 10^{-22}$ eV, and on the SIDM kinetic temperature $T_X\gtrsim 0.4\, (\gtrsim 0.3)$ keV, which are competitive yet independent with respect to current literature constraints from a variety of other probes.

Our analysis highlights the relevance of upcoming ultra-faint galaxy surveys in the (pre)reionization era via JWST (see \cite{Park20,Labbe21,Robertson22}) as a direct probe both of the astrophysics of galaxy formation at small scales, and of the microscopic nature of the elusive dark matter particles.

\vspace{-3pt}

\funding{We thank the three anonymous referees for useful comments and suggestions. A.L. acknowledges funding from the EU H2020-MSCA-ITN-2019 project 860744 \textit{BiD4BESt: Big Data applications for black hole Evolution STudies} and from the PRIN MIUR 2017 prot. 20173ML3WW, \textit{Opening the ALMA window on the cosmic evolution of gas, stars, and supermassive black holes}.}

\acknowledgments{We thank L. Boco, C. Baccigalupi, and P. Salucci for stimulating discussions.}

\conflictsofinterest{The authors declare no conflict of interest.}

\begin{adjustwidth}{-\extralength}{0cm}

\reftitle{References}
\printendnotes[custom]

\end{adjustwidth}

\end{document}